\documentclass[twocolumn,english,aps]{revtex4}
\usepackage[LGR,T1]{fontenc}
\usepackage[latin9]{inputenc}
\usepackage{amsmath}
\usepackage{graphicx}
\usepackage{esint}

\makeatletter

\DeclareRobustCommand{\greektext}{%
  \fontencoding{LGR}\selectfont\def\encodingdefault{LGR}}
\DeclareRobustCommand{\textgreek}[1]{\leavevmode{\greektext #1}}
\DeclareFontEncoding{LGR}{}{}
\DeclareTextSymbol{\~}{LGR}{126}

\@ifundefined{textcolor}{}
{%
 \definecolor{BLACK}{gray}{0}
 \definecolor{WHITE}{gray}{1}
 \definecolor{RED}{rgb}{1,0,0}
 \definecolor{GREEN}{rgb}{0,1,0}
 \definecolor{BLUE}{rgb}{0,0,1}
 \definecolor{CYAN}{cmyk}{1,0,0,0}
 \definecolor{MAGENTA}{cmyk}{0,1,0,0}
 \definecolor{YELLOW}{cmyk}{0,0,1,0}
}

\makeatother

\usepackage{babel}
\begin{document}

\title{The role of curvature anisotropy in the ordering of spheres on an
ellipsoid}

\author{Christopher J. Burke}

\affiliation{Tufts University, Department of Physics and Astronomy, Center for
Nanoscopic Physics, 5 Colby Street, Medford, Massachusetts, 02155}

\author{Badel L. Mbanga}

\affiliation{Tufts University, Department of Physics and Astronomy, Center for
Nanoscopic Physics, 5 Colby Street, Medford, Massachusetts, 02155}

\author{Zengyi Wei}

\affiliation{University of New South Wales, Department of Chemical Engineering}

\author{Patrick T. Spicer}

\affiliation{University of New South Wales, Department of Chemical Engineering}

\author{Timothy J. Atherton}

\email{timothy.atherton@tufts.edu}

\affiliation{Tufts University, Department of Physics and Astronomy, Center for
Nanoscopic Physics, 5 Colby Street, Medford, Massachusetts, 02155}
\begin{abstract}

Non-spherical emulsion droplets can be stabilized by densely packed
colloidal particles adsorbed at their surface. In order to understand
the microstructure of these surface packings, the ordering of hard
spheres on ellipsoidal surfaces is determined through large scale
computer simulations. Defects in the packing are shown generically
to occur most often in regions of strong curvature; however, the relationship
between defects and curvature is nontrivial, and the distribution
of defects shows secondary maxima for ellipsoids of sufficiently high
aspect ratio. As with packings on spherical surfaces, additional defects
beyond those required by topology are observed as chains or ``scars''.
The transition point, however, is found to be softened by the anisotropic
curvature which also partially orients the scars. A rich library of
symmetric commensurate packings are identified for low particle number.
We verify experimentally that ellipsoidal droplets of varying aspect
ratio can be arrested by surface-adsorbed colloids.
\end{abstract}
\maketitle

\subsubsection*{Significance statement}

Emulsions, combinations of immiscible fluids such as oil and water,
comprise a variety of commercially relevant systems, from ice cream
to cosmetics. Individual droplets in these emulsions can be stabilized
by the presence of colloidal particles and nonspherical droplets can
be produced by a mechanism of arrested relaxation, where particles
adsorbed at the droplet interface become crowded and obstruct further
evolution of the surface toward the spherical ground state. The particles
tend to have a high degree of crystalline order, but the curvature
of the surface frustrates this order and introduces defects. In this
paper we study the defect structures that form in packings of hard
particles on ellipsoidal surfaces, providing insight into the microstructure
and stability of arrested systems.

\section{Introduction}

\begin{figure}
\begin{centering}
\includegraphics{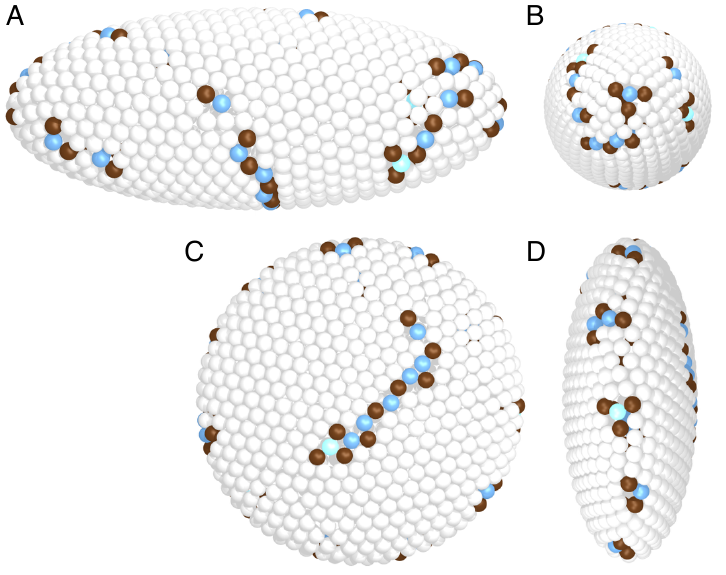}
\par\end{centering}

\centering{}\protect\caption{\label{fig:packingExample}Sample packing of $N=800$ particles on
a prolate ellipsoid of aspect ratio $2.6$. (\textit{A}) Side and
(\textit{B}) end views are shown; corresponding plots are shown for
an oblate ellipsoid of aspect ratio $2.6$ (\textit{C}) from the top
and (\textit{D}) around the rim. Particles are colored by coordination
number as computed from the Delaunay triangulation of the centroids
--- 5: brown; 6: white; 7: blue; 8: light blue. }
\end{figure}
Emulsions---mixtures of two immiscible fluids---are ubiquitous systems
with many applications in the food, oil, and cosmetics industries.
At the microscopic level, an emulsion consists of droplets of one
fluid embedded in a host fluid; the droplets are held in an equilibrium
spherical shape by the interfacial tension between the two fluids.
Emulsions with anisotropic droplets are of interest because for some
applications, e.g. particle filtering in porous media\cite{Weiss1995},
performance is improved with increasing aspect ratio. Anisotropic
particles are also known to be more easily absorbed by cells, thus
being effective as drug delivery systems\cite{Champion2006,Gratton2008}.
Additionally, ellipsoids fill space more efficiently than spheres\cite{Donev2004a},
and through chemical functionalization, are a valuable component in
the nano-architecture of hierarchical structures\cite{Nelson2002}. 

A mechanism for sculpting stable shaped droplets exists in Pickering
emulsions, where the constituent droplets are stabilized by colloidal
particles adsorbed at the interface\cite{Pickering1907}. The particles
are strongly bound to the surface because they reduce the interfacial
tension between the two immiscible phases\cite{Binks2002}. Non-spherical
shapes can be produced by a sequence of deformation, adsorption, relaxation
and arrest as follows: Following an initial deformation, for example
by an applied electric field\cite{Cui2013} or by the coalescence
of two droplets\cite{Pawar2011}; during this process additional particles
may become adsorbed on the interface from the host fluid. The droplet
then relaxes towards the equilibrium spherical shape, reducing the
surface area and causing the particles to become more densely packed.
If the surface coverage of colloids is sufficiently high, they will
become crowded and arrest the shape evolution of the droplet before
a spherical shape is reached\cite{Cheng2009,Pawar2011}. 

The purpose of this paper is to identify the role that the anisotropic
curvature present in an ellipsoid plays on the ordering of the particles.
We assume the particles interact purely through volume exclusion.
The quality of the packing of the final state, measured globally by
coverage fraction as well as locally by coordination number, depends
on this ratio: As $\tau_{r}/\tau_{d}\to0$, the particles are unable
to rearrange themselves significantly and may get trapped in a glassy
state, while for $\tau_{r}/\tau_{d}\to\infty$, the relaxation proceeds
slowly and the situation resembles a classical sphere packing problem.
It is this latter quasi-static limit of the relaxation process that
we shall examine in this work. 

Since the colloids are confined to a 2D surface, the arrested states
tend to be quite crystalline as has been shown for spherical droplets
or \emph{colloidosomes}\cite{Dinsmore2002}. These structures should,
therefore, exhibit properties similar to 2D elastic crystalline membranes\cite{Seung1988,Perez-Garrido1997,Bowick2000,Bowick2001,Bausch2003,Einert2005,Vitelli2006,Giomi2007,Giomi2008,Irvine2010,Bendito2013}.
The presence of curvature frustrates the crystalline order and induces
defects: particles which have more or fewer than six neighbors, and
whose deviation from six-fold order can be quantified as a topological
charge: particles with coordination number lower than six have positive
charge and \emph{vice versa}. Lone defects of positive or negative
charge are known as \emph{disclinations}. The topology of the droplet
surface will determine the net defect charge, which is 12 for a spherical
topology\cite{Hilton1996}. Furthermore, there is a coupling of defects
to the Gaussian curvature $K$. Because droplets with non-spherical
geometries possess a variation in Gaussian curvature along their surface,
the defects should be non-uniformly distributed as theoretical studies
have predicted\cite{Giomi2007,Giomi2008,Vitelli2006}.

In addition to the minimal number of defects required by topology,
pairs of positive and negative defects called dislocations can occur.
Droplets with a large system size , i.e. where the ratio $R/r$ of
the droplet size $R$ to the particle size $r$ is large enough, exhibit
chains of defects known as scars\cite{Bowick2000,Bausch2003}. For
spherical droplets, a transition has been shown: $R/r$ is below a
critical value only isolated defects occur. Above this ratio, scars
appear and increase in length with $R/r$\cite{Bausch2003}.

For surfaces of nonuniform curvature, the placement of the defects
is an interesting question. The theory of curved elastic crystalline
membranes\cite{Bowick2000} predicts that defects and Gaussian curvature
act as source terms in a biharmonic equation,
\begin{equation}
\nabla^{4}\chi(\vec{x})=\rho(\vec{x})-K(\vec{x}),\label{eq:biharmonic}
\end{equation}
where $\chi$ is a stress function and $\rho$ is the defect charge
density (a sum of point charges). The energy of such a system is,
\begin{equation}
U=\int_{S}dA\chi(\vec{x})(\rho(\vec{x})-K(\vec{x})),\label{eq:energy}
\end{equation}
which must be minimized with respect to defect number and defect position,
with total defect charge conserved according to the surface topology.
While this suggests that defect charges will be attracted to areas
of like-signed curvature in order to minimize the source term, the
fact that these systems are governed by a biharmonic equation suggests
that the coupling of defects to curvature is nontrivial. This is in
contrast to simpler analogues, for example electrostatics, governed
by a Poisson equation.

There are two important differences between an elastic crystalline
membrane and a 2D arrested hard sphere system. First, in the hard
sphere limit, the in-plane elastic constants of a hard sphere system
are infinite. Second, arrested hard sphere systems are not able to
explore their full phase space, and so one does not expect to find
them in the optimally packed ground state. For these reasons, it should
not be expected that a 2D hard sphere system is exactly described
by the theory of 2D crystalline membranes, but due to the highly crystalline
order that is exhibited, the behavior should be qualitatively similar.
Additionally, the relative wetting properties of the two fluids may
induce a contact angle, leading to inter-particle interactions that
may modify the ordering\cite{Kralchevsky2000}.

In other systems in the packing limit, e.g. viral capsids\cite{Roos2010}
and small clusters of colloids\cite{Manoharan2003}, configurations
with a high degree of symmetry are typically observed for certain
special numbers of particles. Experimentally, these tend to be stable,
and so the identification of possible symmetric packings may serve
as a guide towards stable self-assembled micro-structures. We therefore
examine the packings systematically by aspect ratio $a$ and particle
number $N$ to identify the symmetric configurations.

In order to explore the role of surface anisotropy on the ordering
of packed particles, we present the results of simulations of hard
spheres packed onto ellipsoidal surfaces using an inflation algorithm.
Sample results are shown in fig. \ref{fig:packingExample}. We investigate
the effect of aspect ratio and particle number on the average distribution
of defects on our surfaces and the structure of the defects themselves.
We also identify highly symmetric configurations. Experimentally,
we demonstrate that ellipsoidal droplets can be stabilized by surface-adsorbed
colloids, and we compare the spatial distribution of defects in the
experiments and simulations. Details of the model and simulations
are presented in \textit{Methods}.

\section{Results and Discussion}

We employ an inflation packing algorithm in order to generate packings
of spheres on ellipsoidal surfaces. The centroids of $N$ equal sized
spheres are bound to a fixed ellipsoidal surface, either prolate or
oblate, of aspect ratio $a$. The particles have hard-sphere interactions
and diffuse as the particle radius is slowly incremented, until further
inflation is precluded. Further details of the algorithm are given
in \textit{Methods}.

Two sets of data were generated from which we obtained our results.
One data set was used for studying the curvature-defect coupling and
scar length, which consisted of packings with aspect ratio varying
from 1.2 to 4.0 in increments of 0.2 (for both the prolate and oblate
cases: we consider the aspect ratio to be the ratio of the semi-major
to semi-minor axis.) The particle number was varied from 10 to 800
in increments of 10. Additional prolate packings were generated to
study scar orientation, from aspect ratio 4.2 to 8.0 in increments
of 0.2, from particle number 710 to 800 in increments of 10. 50 configurations
were generated for each pair of parameters. The second data set was
used for studying symmetry, where we are interested in lower particle
numbers and a more fine-grained search of the parameter space. This
data set consisted of packings with aspect ratio varying from 1.1
to 4.0 in increments of 0.1, and particle number varying from 3 to
200 in increments of unity. 80 configurations were generated for each
pair of parameters.

\subsection{Defect Distribution\label{sub:Defect-distribution}}

\begin{figure}
\includegraphics{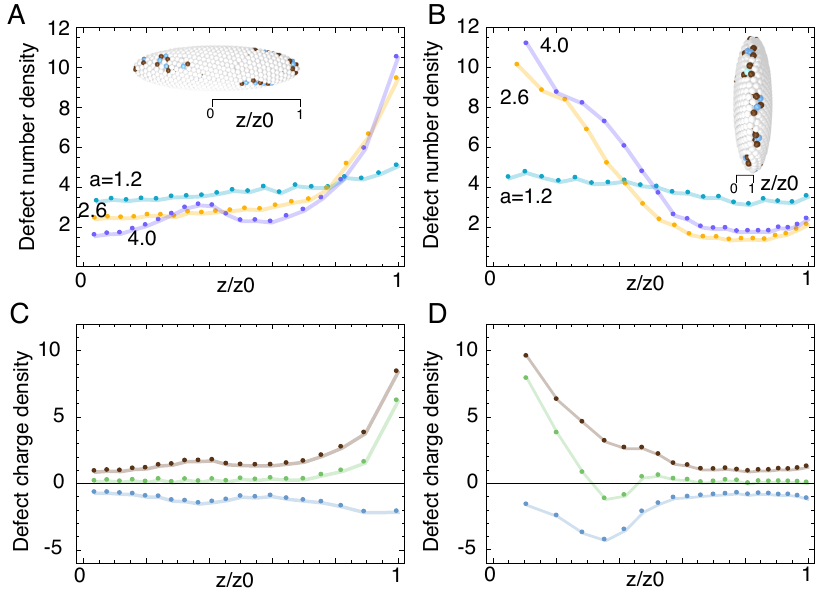}

\centering{}\protect\caption{\label{fig:defectDist}Defect number density for (\textit{A}) prolate
and (\textit{B}) oblate ellipsoids of varying aspect ratio: blue is
1.2; yellow 2.6; purple 4.0. Note the small secondary peak near $z/z_{0}=0.4$
at $a=4$ in the prolate case. Example configurations of $a=4$ are
shown as insets. Defect charge density is shown for (\textit{C}) prolate
and (\textit{D}) oblate ellipsoids of $a=4$. The green points represent
the net charge density, and the brown and blue points represent the
density of positive and negative defects, respectively. The secondary
peak in (\textit{A}) is also visible in the positive and negative
charge densities in (\textit{C}). In (\textit{D}), there is a net
negative defect charge density near $z/z_{0}=0.4$, despite the Gaussian
curvature being positive. Lines are guides to the eye.}
\end{figure}

We first examined the distribution of the defects as a function of
the aspect ratio. Defect locations were determined by assigning a
defect charge $q=6-c$ to each particle, where $c$ was the coordination
number determined from the Delaunay triangulation of the particle
positions (see \textit{Methods}). The surface was partitioned into
equal-area axisymmetric regions and the number of defects in each
region counted. Each segment has a different average Gaussian curvature
with regions near the poles having larger curvature for prolate and
the reverse for oblate ellipsoids. In fig. \ref{fig:defectDist}\textit{A}
for prolates and fig. \ref{fig:defectDist}\textit{B} for oblates,
the defect number density is shown as a function of the axial position
$z/z_{0}$ averaged over the ensemble of simulations at fixed aspect
ratio and particle numbers ranging from $710<N<800$. Generically,
it is apparent that defect number density increases with the Gaussian
curvature, as expected. For prolate ellipsoids at low aspect ratio,
the defect number density increases monotonically with respect to
$K$. At higher aspect ratios, there is a small secondary peak in
segments with low Gaussian curvature. We verified this occurs for
other ranges of particle numbers $N>210$.

In order to understand this, we plot separate defect charge densities
for positive and negative defects in fig. \ref{fig:defectDist}\textit{C},
as well as the net defect charge density. The anomalous peak is apparent
in both the separate positive and negative defect charge densities,
but not in the net defect charge density, indicating that the excess
defects are taking the form of neutral dislocations or scars.

In fig. \ref{fig:defectDist}\textit{B}, we see that for oblate ellipsoids,
the defect density again increases near the more highly curved regions.
Fig. \ref{fig:defectDist}\textit{D} reveals, however, that the coupling
between defect charge and curvature is again complicated: while there
is a peak in positive defects at the highly positively curved edge
of the surface, there is a high density of negative defects surrounding
this, and the net defect charge density is actually negative near
$z/z_{0}=0.4$.

These results display a non-trivial interaction between defects and
curvature. While the regions of highest Gaussian curvature contain
the highest density of defects, the defect density is not a simple
monotonic function of Gaussian curvature. This is apparent in the
defect number density in the prolate case, and in the defect charge
density in the oblate case. The fact that the defect charge density
can be negative in regions of positive Gaussian curvature is especially
surprising. However, this is not necessarily inconsistent with eqs.
\ref{eq:biharmonic} and \ref{eq:energy}, which imply complex defect
behavior. Further investigation is warranted to confirm whether the
continuum elastic theory gives results similar to the hard sphere
packings here.

\subsection{Scar Orientation\label{sub:Defect-orientation}}

\begin{figure}
\raggedright{}\includegraphics[width=3.3in]{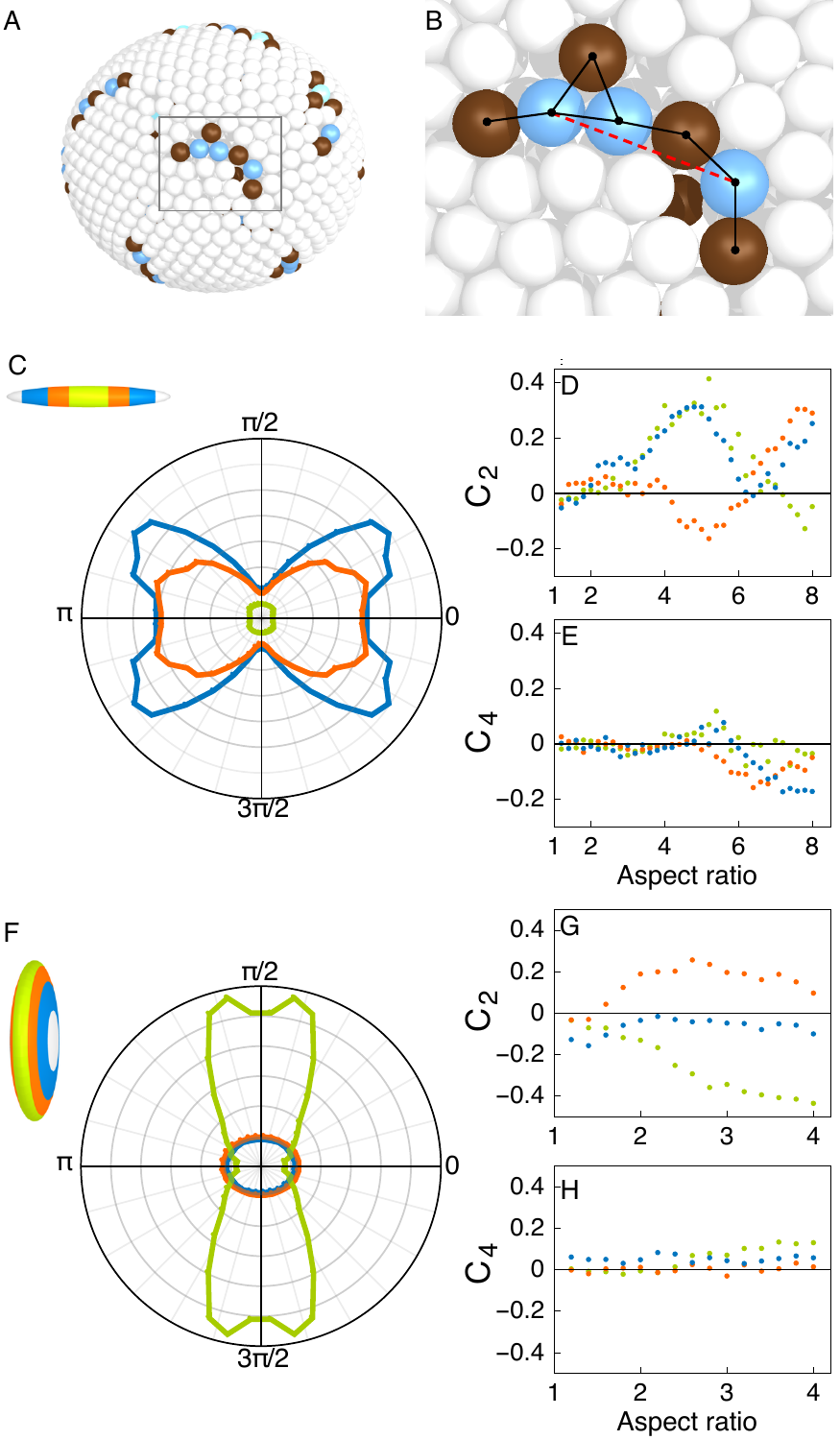}\protect\caption{\label{fig:orientation}Orientation of the scars relative to the curvature
anisotropy. (\textit{A}) A configuration with a typical scar. (\textit{B})
Close-up of the scar. Black lines show edges in a graph comprising
the scar. The red dashed line shows a chain of length 3. Results are
shown for (\textit{C}-\textit{E}) prolate and (\textit{F}-\textit{H})
oblate ellipsoids. The $C_{2}$ (\textit{D}, \textit{G}) and $C_{4}$
(\textit{E}, \textit{H}) order parameters for prolate and oblate ellipsoids,
respectively, are plotted as a function of aspect ratio for different
regions along the symmetry axis of the ellipsoid: green corresponds
to the center, orange to the mid-region, and blue to the ends. (\textit{C})
and (\textit{F}) show the ODF of chains in the center, mid-regions,
and ends of the ellipsoid, respectively, for prolate ellipsoids of
aspect ratio 8 in (\textit{C}) and oblate ellipsoids of aspect ratio
4 in (\textit{F}). Insets of (\textit{C}) and (\textit{F}) illustrate
the regions used for spatial binning.}
\end{figure}

We next determined whether the scars are oriented by the curvature
anisotropy of the surface. To do so, we consider a local scar orientational
distribution function (ODF) $f(\alpha)$ where the angle $\alpha$
is measured locally in the tangent plane relative to the uniaxial
axis of the ellipsoid. The ODF may be expanded as a Fourier series,
\begin{equation}
f(\alpha)=\sum_{n}C_{n}\cos(n\alpha).
\end{equation}

The average value of the first two non-zero coefficients, $C_{2}=\langle cos(2\alpha)\rangle$
and $C_{4}=\langle cos(4\alpha)\rangle$, were calculated for our
ensemble of packings. These quantities are order parameters for orientational
order as they vanish if the scars align isotropically with the curvature.
$C_{2}$ quantifies nematic order, i.e. uniaxial orientational order
and $C_{4}$ quantifies quadrupolar order.

To determine the scar orientation, we studied contiguous chains of
defects as shown in fig. \ref{fig:orientation}\textit{A} and \textit{B}.
Given a packing and its Delaunay triangulation, the neighboring defects
around each defect are identified. These adjacent pairs become the
edges of graphs of contiguous defects. Two defects are identified
as the ends of a \textit{chain} of length $l$ if they are within
a connected graph of defects and the shortest path between them contains
$l$ edges. Once a chain of length $l$ is identified, its orientation
relative to the local principal directions\textemdash i.e. the polar
and azimuthal tangent vectors $\vec{t}_{\theta}$ and $\vec{t}_{\phi}$,
respectively (see eq. \ref{eq:surface-1} in the Appendix for the
parametrization of the surface)\textemdash is calculated thus: given
a pair of chain endpoints, their separation vector is projected onto
the surface at each endpoint, giving components along $\vec{t}_{\theta}$
and $\vec{t}_{\phi}$. These components are then averaged between
the endpoints, and the angle $\alpha$ that the resulting vector makes
with $\vec{t}_{\theta}$ is recorded as the orientation of the chain.
The $z$-component of the midpoint of each chain is recorded as its
position and is used to examine how the coupling varies across the
surface.

The analysis was applied to an ensemble of simulation results as follows:
For a given aspect ratio, the orientations of all chains of length
$l$ are collected across simulations with $N\in[710,800]$ in increments
of $\Delta N$=10 (with 50 results at each $N$ resulting in 500 simulations).
Order parameters $C_{2}$ and $C_{4}$ are then calculated from this
ensemble. Because the curvature anisotropy varies with the $z$-coordinate
along the surface, results can be divided according to their position.
In our analysis, we exclude scars in the regions near the poles which
make up 10\% of the surface area as here the curvature tensor is degenerate
and the alignment is undefined. The rest of the surface is broken
into six equal-area, azimuthally symmetric regions, as illustrated
in the insets of fig. \ref{fig:orientation} \textit{C} and \textit{F},
and data from symmetric regions on opposite halves of the ellipsoid
are combined. A chain length of $l$=3 was used as this is long enough
to capture scar behavior while having enough chains for statistical
purposes. Shorter chain lengths show a weaker tendency to orient.

The behavior exhibited by prolate ellipsoids is rather complicated,
as seen in the plots of order parameter versus aspect ratio in fig.
\ref{fig:orientation}\textsl{ D} and \textsl{E}. In the center region
near the equator, scars are nematic along the $\vec{t}_{\theta}$
direction between aspect ratio 3.6 and 6. At higher aspect ratio this
center region is very flat, leading to fewer scars, and so any orientational
order is insignificant. In the mid-regions between the equator and
poles, scars become nematic along the $\vec{t}_{\phi}$ direction
at aspect ratio 4.4, and then transition to nematic along the $\vec{t}_{\theta}$
direction at aspect ratio 6.4. Scars near the poles show nematic order
along $\vec{t}_{\theta}$ above aspect ratio 2, although this order
peaks near aspect ratio 5, then drops to $C_{2}$=0 at aspect ratio
6.4 before increasing again. Interestingly, scars on highly prolate
ellipsoids can also show $C_{4}$ order. This appears in the mid regions
above aspect ratio 5.2, and in the end regions above aspect ratio
6.

The chain ODFs for prolate ellipsoids of aspect ratio $a=8$ in fig.
\ref{fig:orientation}\textsl{C} illustrate the trends that appear
at high aspect ratio. It is apparent from the green curve that that
there are few chains in the relatively flat center of the ellipsoid.
The orange curve shows a high degree of nematic order directed along
the polar direction in the mid-region, and the blue curve for the
ends shows nematic order along the polar direction, as well as a peak
between the directions of principal curvature, which is indicative
of negative $C_{4}$ order.

The case of scar orientation on oblate ellipsoids is more straightforward.
The order parameters are plotted as a function of aspect ratio for
different azimuthally symmetric regions across the surface, in fig.
\ref{fig:orientation} \textsl{G} and \textsl{H}. Scars at the equator
exhibit a high degree of nematic order in the $\vec{t}_{\phi}$ direction,
which increases linearly with aspect ratio up to $a=4$. This is unsurprising,
because the curvature on highly oblate ellipsoids is localized to
a nearly one-dimensional region around the equator of the ellipsoid,
and so one expects the scars to form there, aligned along the equator.
There is also a small degree of $C_{4}$ ordering. In the regions
midway between the equator and poles, there is a weak coupling of
scars along the $\vec{t}_{\theta}$ direction. These trends are illustrated
for $a=4$ in fig. \ref{fig:orientation}\textsl{F}:the green curve
for the edges displays a peak near the azimuthal direction, whereas
the orange and blue curves show that there are fewer chains without
much order in the flatter regions.

While the scar orientation results for the oblate case are easily
understood, the ordering of the scar orientation on prolate ellipsoids
is far more complicated. The orientation varies greatly depending
on chain position and ellipsoid aspect ratio. Especially surprising
is the emergence of $C_{4}$ ordering, which corresponds to a tendency
for chains to align in a direction intermediate to the directions
of principal curvatures.

\subsection{Scar Transition\label{sub:Scar-transition}}

\begin{figure}
\includegraphics[width=3.4in]{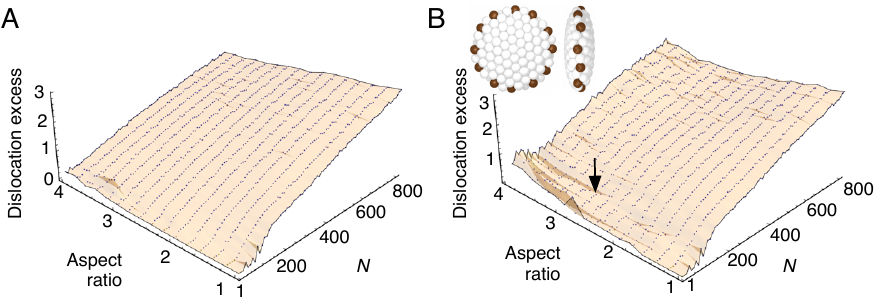}

\centering{}\protect\caption{\label{fig:scarTransitionProlate}The number of excess dislocation
defects per scar on (\textsl{A}) prolate ellipsoids and (\textsl{B})
oblate ellipsoids. For low aspect ratio near 1, there is a clear scar
transition, which is not present at aspect ratios far from 1. The
inset in (\textsl{B}) shows a highly commensurate oblate packing with
$N=140$ and $a=2.6$. Note that data for oblate ellipsoids with $N=10$,
$a\ge2.0$ and $N=20$, $a\ge3.0$ has been excluded. .}
\end{figure}

As is well known from previous work\cite{Bowick2000,Bausch2003},
packings of spheres on spherical surfaces exhibit a transition: For
low particle numbers, only the twelve defects required by topology
are present; above a critical particle number $N_{c}$, it is favorable
for larger defect structures to occur, typically chains of scars extending
from a core disclination. Increasing $N$ above $N_{c}$ leads to
a monotonic increase in average scar length. 

From our simulation results of packings with $10\le N\le800$, we
calculated the average number of excess dislocations per topologically
required disclination for each $(a,N)$. Defects were weighted in
the analysis by the absolute value of their charge. Given that there
are two disclinations per dislocation, and 12 core disclinations,
the number of excess dislocations per scar is calculated thus,
\begin{equation}
n_{d}=\frac{1}{2}\left(\frac{\sum_{i}\left|q_{i}\right|}{12}-1\right),
\end{equation}
where the sum is taken over all defects. This quantity captures the
same information as the scar length but is easier to calculate, as
individual scars are often not well defined. 

Results of the analysis are displayed in fig. \ref{fig:scarTransitionProlate}.
Prolate ellipsoids {[}fig. \ref{fig:scarTransitionProlate}\textsl{A}{]}
show the experimentally observed behavior for low aspect ratio: for
$N<100$ particles there are few excess defects, but at higher particle
numbers there is a roughly linear increase in the number of excess
defects. As aspect ratio increases, however, the transition is softened
such that there is a smooth increase in excess defects with $N$.
This is reminiscent of how applied fields soften phase transitions\cite{Landau};
here the anisotropy of the curvature seems to play a similar role. 

The oblate packings show the same trends {[}fig. \ref{fig:scarTransitionProlate}\textsl{B}{]}.
There is, however, an additional feature that stands out. At $N=140$,
$a>2$, there is a set of nearly scar-free configurations. This is
due to commensurability as the particle number and surface geometry
for these cases are compatible with a highly symmetric packing with
only the minimally required defects, as seen in the inset of fig.
\ref{fig:scarTransitionProlate}\textsl{B}. Similar commensurability
issues occur in other systems, e.g. sphere packings on cylinders\cite{Wood2013}.

\begin{figure}
\includegraphics[width=3.4in]{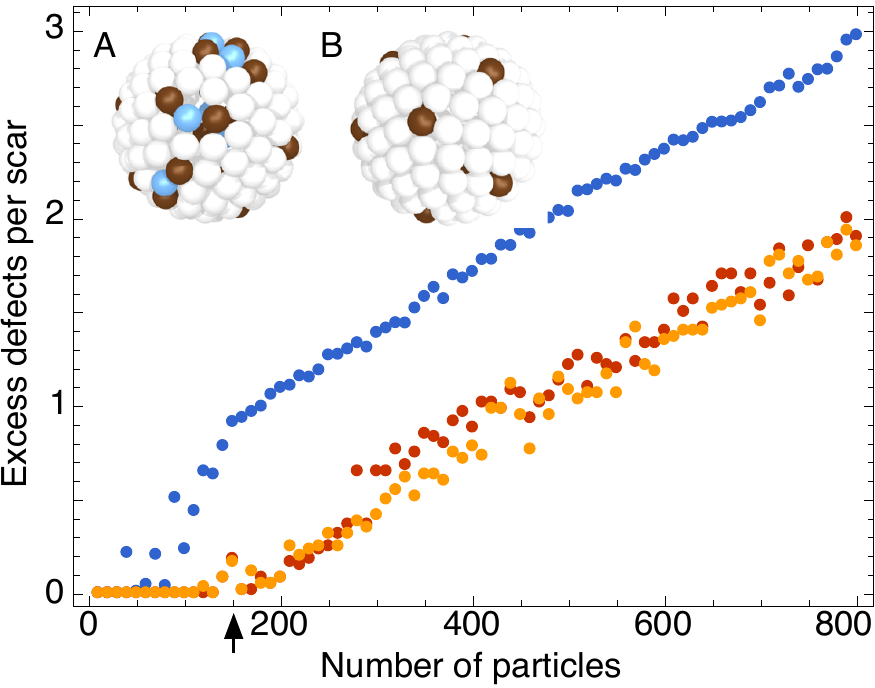}

\protect\caption{\label{fig:softparticles}Excess dislocations per scar as a function
of particle number for hard (blue) and soft $V=1/d$ (orange) and
$V=1/d^{6}$ (red) interactions. Inset (\textsl{A}) is a hard particle
packing and inset (\textsl{B}) is a soft particle packing. The arrow
indicates the particle number of the inset packings.}
\end{figure}

A striking difference between these results and those from a previous
study is that here, for hard particles, the transition occurs at a
lower particle number; in ref. \cite{Bausch2003} it was seen at $N_{c}\approx400$
using colloidal particles with a soft repulsive interaction. We therefore
performed simulations (see \textit{Methods}) using two different potentials,
$V=d^{-1}$ and $V=d^{-6}$ (where $d$ is the interparticle separation),
the results of which are shown in fig. \ref{fig:softparticles}. For
soft particle packings, we take the average scar length of the five
lowest energy configurations obtained out of an ensemble of 50. For
the hard spheres, $N_{c}\approx80$, while for the two soft potentials
the transition occurs around $N_{c}\approx200$ (which appears to
be within the uncertainty of the result presented in ref. \cite{Bausch2003}).
The defect number increases at the same rate with respect to particle
number for both soft potentials. This supports the conclusion in ref.
\cite{Bausch2003} that, for soft particles, the scar transition does
not depend on the specific form of the particle potential. For hard
particles we have quantitatively different behavior. Visual inspection
of hard and soft sphere configurations reveals that hard sphere configurations
possess gaps (fig. \ref{fig:softparticles}\textsl{A}). It is rare
to find a lone disclination; it is much more common to find a disclination
attached to one dislocation (i.e. a small 5-7-5 scar) adjacent to
a gap in the packing. This isn't seen in soft particle configurations
(fig. \ref{fig:softparticles}\textsl{B}), as the energy penalty is
too high, rather a particle can be squeezed to fill in the gaps. The
fact that hard particle packings tend to have gaps makes them especially
suitable for chemical functionalization as described in ref. \cite{Nelson2002}.

\subsection{Packing Fraction and Symmetry}

We now turn to how the packing fraction varies with respect to particle
number and ellipsoid aspect ratio. To simplify the calculation we
make the approximation, valid for large $N$, that the area covered
by a particle is its projection onto a flat 2D surface, 
\begin{equation}
\phi=\frac{N\pi r^{2}}{A},\label{eq:packingfraction-1-1}
\end{equation}
where $A$ is the area of the underlying surface. We checked the validity
of this estimate by numerically integrating the area of intersection
between the surface and the spheres on oblate surfaces of aspect ratio
4.0, and found that the difference between our estimate and the true
value is very small: using the projected area underestimates the packing
fraction by approximately 1\% for packings with $N=100$ and 0.1\%
for packings with $N=800$. 

For large $N$, the packing fraction increases slightly with aspect
ratio. This is because for large $a$ the curvature\textemdash and
hence the defects\textemdash are mainly localized to the poles on
prolate surfaces or the equator on oblate surfaces and so more of
the surface can be covered by the planar hexagonal packing, consistent
with the results of the above subsections on the \textit{Defect Distribution}
and \textit{Scar Transition}. For low $N$, the opposite tends to
be true; the packing fraction decreases with aspect ratio. However,
the trend is more complex and the packing fraction is sensitive to
both $N$ and $a$ at low $N$. Visual inspection of these configurations
reveals that for specific combinations of $N$ and $a$, the packings
have a high degree of symmetry, suggesting a commensurability effect,
such as that seen in the \textit{Scar Transition} subsection above.

\begin{figure}
\includegraphics[width=3.4in]{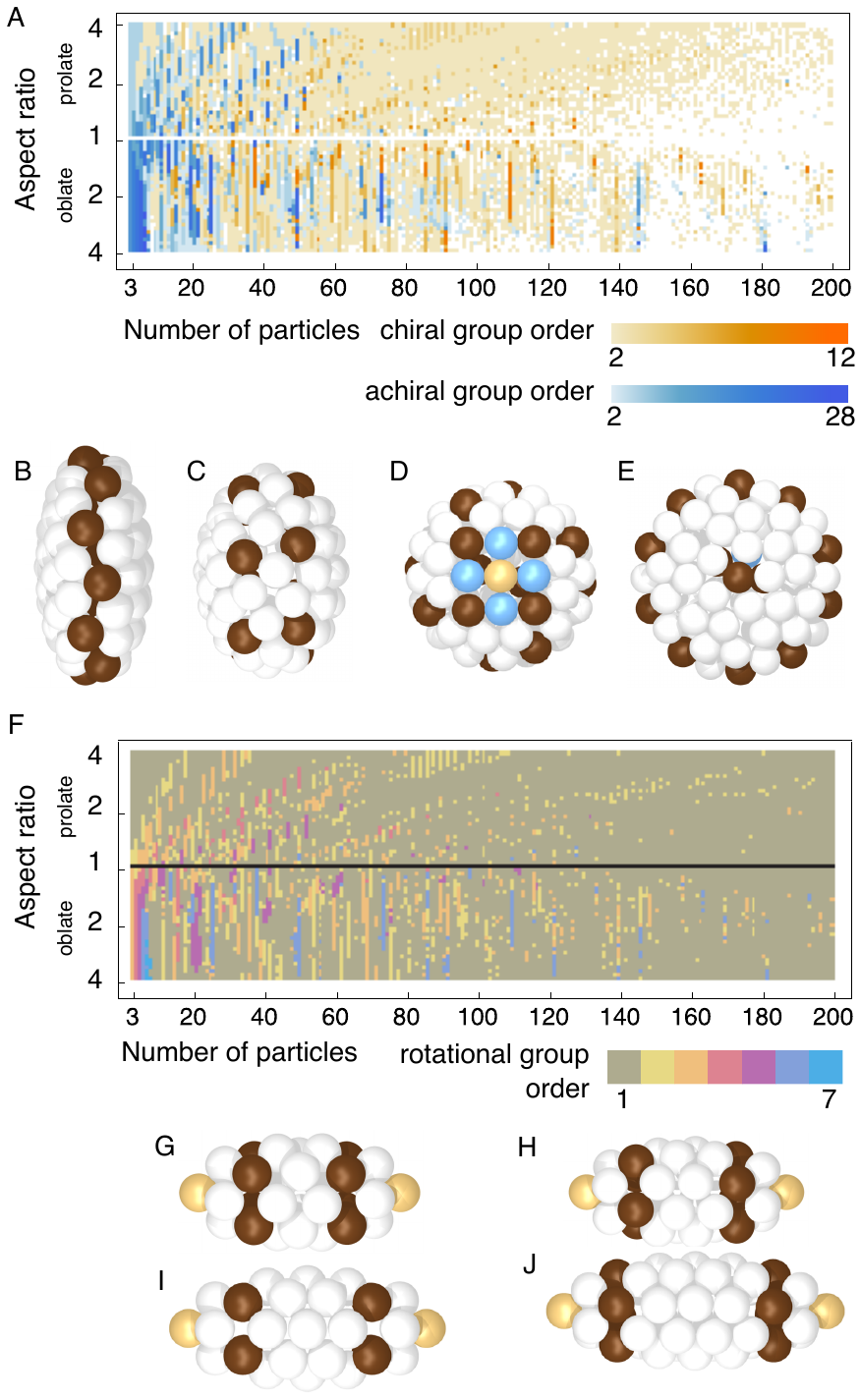}

\protect\caption{\label{fig:The-symmetry-landscape-1}The symmetry landscape for packings
with varying particle number and aspect ratio, using a symmetry norm
cutoff of 0.1. (\textsl{A}) shows the chirality and the order of the
largest symmetry group found. Orange represents chiral packings and
blue represents achiral packings. The boldness of the color corresponds
to the order of the packing's symmetry group as shown in the key.
Note that packings whose only symmetry is the identity are colored white
to distinguish them as being trivially symmetric. Sample packings
are shown: \textsl{B}) a chiral packing with $N=74$, $a=2.5$; \textsl{C})
an achiral packing with $N=74$, $a=1.5$ --- note that (\textsl{B})
and (\textsl{C}) have the same particle number, but show different
chirality for different aspect ratio; \textsl{D}) a packing with fourfold
rotational symmetry with $N=69$, $a=1.4$; \textsl{E}) a packing
with fivefold symmetry $N=76$, $a=2.4$. Light brown particles have
$c=4$. (\textsl{F}) shows the degree of rotational symmetry of each
configuration about its ellipsoidal symmetry axis. Note that for both
(\textsl{A}) and (\textsl{F}), no data is shown for $a=1$ (spheres)
as the spherical symmetry group is not a subgroup of $D_{\infty h}$.
Sample packings are shown for \textsl{G}) $N=30$, $a=2.4$; \textsl{H})
$N=34$, $a=2.5$; \textsl{I}) $N=38$, $a=2.7$; \textsl{J}) $N=46$,
; these packings all occur in the diagonal band of fourfold rotational
symmetry in the top left of (\textsl{F}).}
\end{figure}

To identify these commensurate combinations, we conducted a more thorough
search for symmetric packings using the second data set. An arbitrary
packing must break the ellipsoidal symmetry group of the surface and
hence must belong to some finite subgroup of $D_{\infty h}$; most
packings at high particle number do so trivially, retaining only the
identity element. Defining a suitable inner product $(A,B)$ that
measures the distance between two packings, a packing possesses a
symmetry $\mathcal{C}$ if $(A,\mathcal{C}A)=0$ where $\mathcal{C}$
is a group element of $D_{\infty h}$. The elements $\mathcal{C}$
can be constructed from the group generators: i) an infinitesimal
rotation about the ellipsoid symmetry axis; ii) spatial inversion,
and iii) a rotation by $\pi$ about an axis perpendicular to the symmetry
axis. 

We used a norm $(A,B)$ defined such that, 
\begin{equation}
(A,B)=\sqrt{\frac{1}{N}\sum_{i}^{N}\left(\frac{\text{min}_{j}\left|\vec{a}_{i}-\vec{b}_{j}\right|}{r}\right)^{2}},
\end{equation}
where the $\vec{a}_{i}$ and $\vec{b}_{j}$ are the positions of particles
in packings A and B, respectively: for each particle in $A$, the
closest particle in $B$ is found and the separations between these
pairs are divided by the particle radius. The root mean square of
these normalized separations is then taken as the inner product. From
this, together with the group generators, all symmetries such that
$(A,\mathcal{C}A)\le\epsilon$,a threshold separation were found.
From this catalog of symmetries, for a particular configuration the
appropriate group was determined. From a collection of configurations
with a given $(N,a)$, the most symmetric configuration was chosen
by the following procedure. First, the configurations with the largest
symmetry group were identified. Then, for each of these configurations,
the symmetry group element with the highest symmetry norm was identified.
Finally,the configuration with the minimum highest symmetry norm was
chosen as the most symmetric.

The results of this analysis are displayed in fig. \ref{fig:The-symmetry-landscape-1}\textsl{A}
showing the order and chirality of the symmetry group of the best
packing for each combination of particle number and aspect ratio.
The degree of rotational symmetry for each packing is shown in fig.
\ref{fig:The-symmetry-landscape-1}\textsl{F}. One striking feature
is that, for certain particle numbers, long vertical stripes appear
in the plots representing commensurate aspect ratios for that particle
number. Furthermore, low $N$ favors achiral packing while chiral
packings occur more often for higher particle number. For prolates
the stripes occupy a narrow range of aspect ratio and occur in band-like
sequences described by a straight line $a=mN$ with slope $m$. Each
of these sequences corresponds to a different degree of rotational
symmetry $n_{r}$, and the particle numbers in the sequence are separated
by $n_{r}$. Inspecting the configurations in a single sequence, the
difference between a configuration with $N$ particles and the next
with $N+n_{r}$ particles is that an additional row of $n_{r}$ particles
has been inserted in the space created by the longer aspect ratio.
This is illustrated by a sequence of configurations with fourfold
rotational symmetry in fig. \ref{fig:The-symmetry-landscape-1} \textsl{G}-\textsl{J}. 

For oblate ellipsoids, the symmetric configurations for $N$ particles
occur at a much broader range of aspect ratios and symmetric configurations
are observed at much higher $N$ and tend to have six-fold rotational
symmetry. The reason for this is that the high curvature at the end
of the prolate ellipsoids accommodates $n_{r}$-fold defects at the
poles, and these appear to determine the rotational symmetry for the
entire configuration; for oblates, the poles have low curvature and
promote hexagonal packing, hence causing six-fold rotational symmetry
to be more common. Interestingly, other degrees are present including
$n_{r}=4$ and $n_{r}=5$ and these configurations contain regions
of highly oblique packings (fig. \ref{fig:The-symmetry-landscape-1}
\textsl{D} and \textsl{E}). 

In general, these symmetric packings are notable because they contain
a high degree of hexagonal ordering over much of their surface, with
evenly spaced defects throughout. This high degree of regularity should
provide stability to the packed structure, and reduce the likelihood
of failure from irregularly spaced defects.

\section{Experiment}

\begin{figure}
\includegraphics{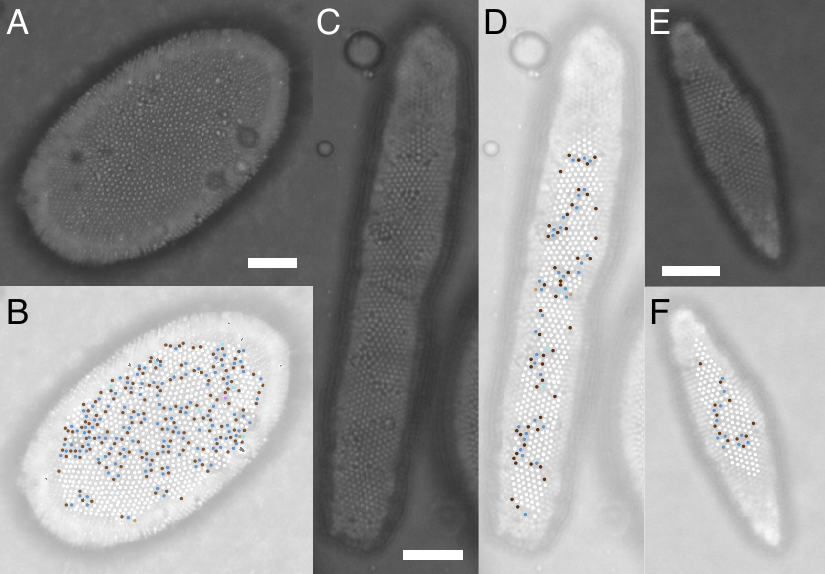}

\protect\caption{\label{fig:experiment}Experimental data for particle-stabilized droplets
of aspect ratio (\textsl{A}, \textsl{B}) 1.6, (\textsl{C, D}) 5.1,
and (\textsl{E}, \textsl{F}) 3.0. Scale bars represent 15$\mu m$.
(\textsl{A}, \textit{C}, \textsl{E}) Microscope images; (\textit{B},
\textit{D}, \textit{F}) Reconstructed particle positions, colored
by coordination number as determined by Delaunay triangulation of
the particle centroids --- 4: light brown, 5: dark brown, 6: white,
7: dark blue, 8: light blue, 9: purple. In general, defects are more
common and are more likely to be found at low-curvature regions of
the droplet in the experiments than in simulations.}
\end{figure}

An experimental realization of ellipsoidal arrested droplets was performed
to confirm the stability of these structures. Ellipsoidal droplets
with arrested interfaces are produced by preparing a Pickering emulsion
and then mixing the emulsion to deform and arrest the droplets in
an elongated shape. Details are given in \textit{Methods}. 

Fig. \ref{fig:experiment} shows several examples of the arrested
droplets observed. Because the curvature of a droplet is significant
across its surface, several focal planes have been combined in the
images in order to study the packing of spheres on the drop surface.
Particle coordinates are determined by finding the local brightness
maxima in the image, recording their coordinates, and correcting for
any unrealistic results via direct comparison with the experimental
images.

Arrest is able to preserve shapes identical to intermediate states
of droplets in an elongation field \cite{Tjahjadi1994}, as seen in
the fig \ref{fig:experiment}\textit{A} and \ref{fig:experiment}\textit{C},
and even shapes resembling sections of such shapes as in the case
of \ref{fig:experiment}\textit{E}. While the dynamical formation
of these shapes was not studied, it is clear that a wide range of
geometries can be formed. We note that the droplet of aspect ratio
5.1 has a spherocylindrical geometry, as opposed to ellipsoidal.

Fig. \ref{fig:experiment} \textit{B}, \textit{D}, and \textit{F}
shows the results of a Delaunay triangulation of the sphere coordinates.
We do not display particles at the boundary of the triangulation,
as they include spurious edge defects identified as a result of the boundary
rather than the ordering of the particles. In each case the arrested
state of the interfacially adsorbed spheres is evident from the visible
regions of crystalline order. Generally, however, the experimental
droplets contain more defects than the simulated packings. In fig.
\ref{fig:experiment}\textit{B} a high degree of hexagonal close-packing
is noted near the ends of the droplet, while the center of the structure
is more disordered with a higher defect density. Three important factors
present in the experiment that are not accounted for in the simulation
may contribute to this. First, the evolution of the surface as it
relaxes will influence particle rearrangement. Different parts of
the surface will grow or shrink at varying rates, affecting where
crowding first occurs. Second, particles adsorbed at an interface
will not act as purely hard spheres. Capillary interactions caused
by the deformation of the surface by the particles will lead to attractive
interactions between particles\cite{Kralchevsky2000}. This may lead
to aggregation of particles during relaxation and is likely to influence
the final ordering of the arrested state. Finally, as discussed in
the introduction, the experimental relaxation does not take place
quasistatically, as is posited by studying the packing limit; it is
highly likely that the particles are arrested in a nonoptimal and
possibly metastable glassy state. A study of the role of these dynamical
influences on the order is in preparation.

\section{Conclusion}

In this paper, we show that defects in the packing of hard spheres
onto an ellipsoidal surface couple nontrivially to the curvature.
For low aspect ratios, the defects occur at regions of high curvature
as predicted by previous studies; additional secondary peaks in the
defect distribution occur in less-curved regions for prolate ellipsoids
of sufficiently high aspect ratio. As previously observed for packings
on a spherical surface, above a critical particle number the defects
take the form of chains or \textquotedblleft scars\textquotedblright{}
rather than isolated defects. This scar transition occurs at a lower
particle number than the previously studied case for soft interparticle
interactions, and is softened by the presence of anisotropic curvature.
The alignment of the scars with the curvature is more complicated:
in flat regions, there is no alignment; in intermediate regions, there
is weak uniaxial alignment with the minimum curvature; in regions
of strong curvature, quadrupolar alignment is seen. We identified
a rich catalog of symmetric configurations from our simulations, each
belonging to a subgroup of the ellipsoidal symmetry group. Plotting
the subgroup order in $(N,a)$ space reveals commensurate surfaces
that promote symmetric packings. Finally, we were able to use the
mechanism of arrest to sculpt ellipsoidal Pickering emulsion droplets
of varying aspect ratio, demonstrating the validity of the fundamental
idea. While careful analysis of these experimental packings reveals
scars as predicted, the defects appear to agglomerate in regions other
than those of strongest curvature, suggesting that dynamical effects
play a significant role in the ordering as well as the geometric effects
studied here.
\begin{acknowledgments}
\emph{TJA and CJB were funded by a Tufts International Research grant
to conduct part of this research at UNSW, Australia. CJB was partly
funded by a Tufts Collaborates! and a Tufts Innovates! grant. ZW was
funded by a UNSW Faculty of Engineering Taste of Research Summer Scholarship.
We would like to thank Marco Caggioni (Procter \& Gamble Co.) for
fruitful discussions about emulsion arrest.}
\end{acknowledgments}

\appendix

\section*{Methods}

\subsection*{Experimental preparation of arrested droplets}

Emulsions are first prepared by mixing 3\% w/w monodisperse 1.5 \textgreek{m}m
diameter precipitated silica particles (Nippon-Shokubai KE-P150) into
hexadecane (Sigma-Aldrich, 99\%)\cite{Pawar2011}. A volume of the
silica-hexadecane dispersion is then emulsified into an equal volume
of deionized water by manual shaking for three minutes. The emulsion
was then aged for 24 hours and inspection revealed a small fraction
of elongated droplets. Imaging of the droplets is carried out on a
Leica DM2500M light microscope using phase contrast optics.

\subsection*{Hard-sphere simulations}

We employ a stochastic inflation packing algorithm inspired by the
Lubachevsky-Stillinger algorithm, which is known to yield packings
of high coverage fraction\cite{Lubachevsky1990}. In each packing
simulation, a fixed ellipsoidal surface, either prolate or oblate,
is chosen with aspect ratio $a$ and the length of the semi-minor
axis is fixed to be unity in dimensionless units. Particles are modeled
as monodisperse hard spheres of radius $r$ that is slowly increased
during the simulation. The number of particles $N$ is specified and
particles are deposited at the start of the simulation by random sequential
adsorption such that the center of each particle is constrained to
lie on the surface of the ellipsoid. Initially, $r$ is such that
the packing fraction is $\phi=0.05$. 

The algorithm proceeds by two kinds of moves: i) Monte Carlo \emph{diffusion
steps} where particles are moved randomly along the surface and ii)
\emph{inflation steps} where the radius of all particles is increased
by $\delta r$. In each diffusion step, $N$ individual Monte Carlo
moves of randomly chosen particles are attempted. The step size is
chosen randomly using a Gaussian distribution, as described below.
Only moves that do not result in overlap are accepted, with overlaps
checked for in the 3D configuration frame.

The moves are performed in the 2D space of conformal surface parameters
$(u,v)$, hence yielding a radially symmetric probability distribution
of moving a certain arclength $s$ in any tangential direction from
the current location. The surface is parametrized as, 
\begin{equation}
x(\theta,\phi)=(x_{0}\sin\theta\cos\phi,x_{0}\sin\theta\sin\phi,z_{0}\cos\theta),\label{eq:surface-1}
\end{equation}
where $x_{0}=1$, $z_{0}=a$ for prolate surfaces and $x_{0}=a$,
$z_{0}=1$ for oblate surfaces. The determinant of the metric is,
\begin{equation}
g(\theta)=\frac{1}{2}x_{0}\sin(\theta)^{2}\left(z_{0}^{2}+x_{0}^{2}+\left(z_{0}^{2}-x_{0}^{2}\right)\cos(2\theta)\right),
\end{equation}
and the conformal parameter $u$ is given by the integral of the conformal
factor,
\begin{equation}
u(\theta)=\int_{\pi/2}^{\theta}\sqrt{g(\theta')}d\theta',
\end{equation}
which can be inverted to find $\theta(u)$. We do an approximate inversion
by calculating $u(\theta)$ for values of $\theta$ from $0$ to $\pi$
in increments of $\pi/100$ and using a high order polynomial least
squares fit on these points, enforcing equality between the fit and
exact values at the endpoints $\theta=0$ and $\theta=\pi$. The conformal
coordinate $v$ is simply $v(\phi)=\phi$.

Given the definitions above, diffusion steps are taken as follows.
An unscaled step size is chosen for each direction, $\Delta u_{o}$
and $\Delta v_{0}$, from a normal distribution with variance 1. These
are scaled by the simulation step size $\sigma$ and by the inverse
of the conformal factor to give step sizes in the $(u,v)$ conformal
space:
\begin{eqnarray}
\Delta u & = & \frac{\sigma\Delta u_{0}}{\sqrt{g(\theta(u))}}\\
\Delta v & = & \frac{\sigma\Delta v_{0}}{\sqrt{g(\theta(u))}}.
\end{eqnarray}
These steps are used to update the previous $u$ and $v$ coordinates
of the particle, which are then transformed to the $\theta$ and $\phi$
coordinates as explained above. Finally, the surface parametrization
eq. \ref{eq:surface-1} is used give the particle coordinates in the
3D configuration space.

Because $\theta$ must have a value between $0$ and $\pi$, we take
the following step if it falls outside this range at any point. If
$u$is greater than $u(0)$ (less than $u(\pi)$), we set $u=2u(0)-u$
($u=2u(\pi)-u$) and $v=\mod(v+\pi,2\pi)$, i.e. we allow the particle
to pass over the coordinate singularity at the poles of the surface.

As the particles diffuse, $\sigma$ is varied in order to more efficiently
explore relevant areas of configuration space (leading to large steps
when the configuration is loosely packed and smaller, more relevant
steps when tightly packed.) The initial value of $\sigma$ scales
with the square root of the ellipsoid surface area $A$, $\sigma_{init}=1\times10^{-4}\sqrt{\frac{A}{4\pi}}$.
After each time step, the fraction of attempted moves that were accepted
is calculated. The length scale $\sigma$ is then decreased by $1\%$
if the acceptance fraction is $<0.5$ and increased by 1\% otherwise;
$\sigma$ is reset after each inflation (described below) to its initial
value. Bounds are imposed such that $1\times10^{-6}<\sigma<1$. Adjusting
$\sigma$ leads to improved performance of the algorithm as the particles
can diffuse more when they are less densely packed and take smaller
steps (which are more likely to be accepted) when they are more densely
packed. We do this as it is known that adaptive algorithms lead to
packings of higher density\cite{Torquato2010}. We emphasize that
in this work the Monte Carlo approach is used as an optimization strategy;
it is not intended to, and indeed cannot, replicate the physical process
by which the structures form since the $\sigma$ updates are non-Markovian.

After $M=100$ diffusion steps, an inflation step is performed where
the particle radius is increased slightly (``inflated'') either
by a specified fixed amount $\Delta r=1\times10^{-5}\sqrt{\frac{A}{4\pi}}$
or by the half of the largest amount allowed that would not result
in the overlap of any pair of particles, whichever is smaller. 

The halting criteria for these simulations is as follows: every $L=100$
inflation steps, the relative change in coverage fraction $\Delta\phi$
is calculated. If this is less than a specified value $\Delta\phi_{tol}=10^{-4}$
then the simulation halts.

\subsection*{Soft particle simulations}

In order to compare our results regarding scar formation in hard particle
packings to other work involving particles with soft interactions,
we performed a set of simulations using a modified Monte Carlo algorithm
which incorporates a soft interparticle potential. In order to test
potentials of different softness, the interparticle potentials are
set as either $U_{int}=d^{-1}$ or $U_{int}=d^{-6}$ (where $d$ is
the center-to-center distance between particles). The particles diffuse
similarly to the hard particle simulation with two differences: the
average step size $\sigma$ is constant for all moves, and moves are
accepted or rejected using a Metropolis scheme\cite{numerical_recipes},
with acceptance probability
\begin{equation}
P=\begin{cases}
1 & \Delta U<0\\
\exp(-\Delta U/k_{B}T) & \Delta U>0
\end{cases}
\end{equation}
where $\Delta U$ is the change in the system energy after a single
particle move. The initial temperature is set by using a rough estimate
of what the energy of a single particle in the final configuration
will be assuming six-fold ordering and that nearest neighbor interactions
dominate: $T_{0}=6U_{int}(2r_{est})/k_{B}$, where $r_{est}=\sqrt{A/N}$
is an estimate of the average particle separation. The system is annealed
by multiplying the temperature by 0.99 after every 100 sets of diffusion
moves until $\exp(-\Delta U/k_{B}T)\to1$ within machine precision.
After every 100 sets of diffusion moves, the change in energy is recorded
and the simulation halts once this change in energy is less than $1\times10^{-16}$.

\subsection*{Defect analysis of simulations}

To analyze defects in the simulated configurations, we use a ball-pivoting
algorithm\cite{Bernardini1999} in the mesh-generation software Meshlab
to generate triangulations of the particle centroids. These triangles
are then equiangulated by a custom script to remove narrow triangles.
Edges are flipped in random order and accepted if they improve the
triangulation; this is repeated until a full sweep of the mesh yields
no further improvements. From these optimized triangulations, the
coordination number of each particle is given by the number of particles
to which it is connected.

\end{document}